\documentclass[pra,aps,10pt,superscriptaddress,twocolumn,floatfix]{revtex4-1}
\usepackage{graphicx,color}
\usepackage[nice]{nicefrac}							
\usepackage{amsmath,amssymb,bm}
\usepackage{ulem}

\usepackage[plainpages=false,pdfpagelabels,colorlinks=true,linkcolor=red,urlcolor=blue,citecolor=blue,pdftitle={},pdfauthor={},pdfdisplaydoctitle=true,pdfduplex=DuplexFlipLongEdge]{hyperref}

\definecolor{darkred}{rgb}{0.90,0.2,0.2}
\definecolor{darkgreen}{rgb}{0,0.60,.2}
\definecolor{darkblue}{rgb}{0.1,0.3,1}
\definecolor{grey}{cmyk}{0,0,0,0.25}
\definecolor{orange}{cmyk}{0,0.6,0.8,0}

\begin{document}

\title{Eigenstate thermalization hypothesis and integrals of motion}

\author{Marcin Mierzejewski}
\affiliation{Department of Theoretical Physics, Faculty of Fundamental Problems of Technology, Wroc\l aw University of Science and Technology, 50-370 Wroc\l aw, Poland}
\author{Lev Vidmar}
\affiliation{Department of Theoretical Physics, J. Stefan Institute, SI-1000 Ljubljana, Slovenia}


\begin{abstract}
Even though foundations of the eigenstate thermalization hypothesis (ETH) are based on random matrix theory, physical Hamiltonians and observables substantially differ from random operators.
One of the major challenges is to embed local integrals of motion (LIOMs) within the ETH.
Here we focus on their impact on fluctuations and structure of the diagonal matrix elements of local observables.
We first show that nonvanishing fluctuations entail the presence of LIOMs.
Then we introduce a generic protocol to construct observables, subtracted by their projections on LIOMs as well as products of LIOMs.
The protocol systematically reduces fluctuations and/or the structure of the diagonal matrix elements.
We verify our arguments by numerical results for integrable and nonintegrable models.
\end{abstract}

\maketitle

{\it Introduction.}
Nonequilibrium dynamics of isolated quantum many-body systems  can nowadays be studied both theoretically and experimentally.
In the latter case,  quantum simulators based on, e.g., quantum gases~\cite{Trotzky2012, Kaufman2016, Neill2016, tang_kao_18} provide a platform to address fundamental questions of quantum mechanics, such as whether and how an initial nonequilibrium system reaches a thermal state~\cite{Bloch2008, polkrev, Langen2015, Eisert2015, dalessio_kafri_16, mori_ikeda_18, deutsch_18}.

Rigorous theoretic approaches, ranging from the von Neumann's quantum ergodic theorem~\cite{vonneumann_29, Goldstein2010}  to the random matrix theory~\cite{dalessio_kafri_16}, provide  general understanding of thermalization in isolated quantum systems.
Nevertheless, much less is known about thermalization of few-body and local operators, which are relevant for experiments.
Indeed, it has recently been shown that few-body operators are atypical~\cite{hamazaki_ueda_18} in the sense that they violate predictions of the theories mentioned above.

\begin{figure}
\centering
\includegraphics[width=\columnwidth]{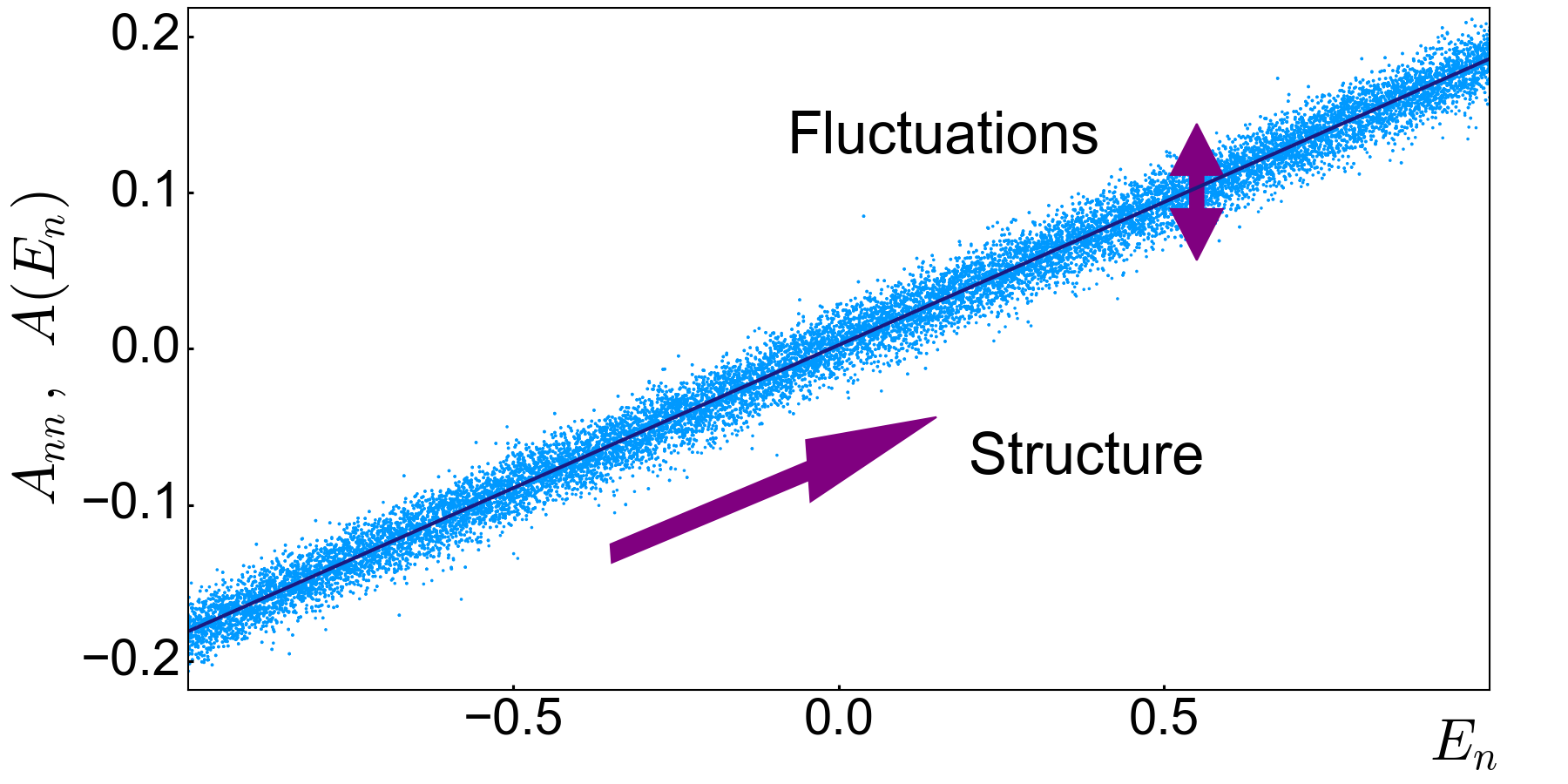}
\caption{
Sketch of the diagonal matrix elements $A_{nn}$ [dots] and the smooth function $A(E_n)$ [line].
Here, $\hat{A}$ is the normalized kinetic energy of a nonintegrable model (\ref{ham1}) at  $L=21$.
Arrows denote two properties studied here: Structure, i.e.,  $A(E_n)\ne$ const, and fluctuations of $A_{nn}$ above $A(E_n)$. 
}
\label{fig0}
\end{figure}

Based on random matrix theory, the eigenstate thermalization hypothesis (ETH) provides a framework to explain thermalization of local observables~\footnote{With {\it local observables} we have in mind an extensive sum of observables with support on ${\cal O}(1)$ sites. In this definition, any local observable is also a few-body observable.} in macroscopic quantum systems~\cite{deutsch91, Srednicki1994, srednicki_99, rigol_dunjko_08, Rigol2012, dalessio_kafri_16}.
While the ETH has not yet been rigorously proved, strong numerical evidence supports validity of the ETH in various generic Hamiltonians with local interactions~\cite{prosen_99, rigol_dunjko_08, pop1, Rigol2009, rigol_santos_10, Neuenhahn2012, steinigeweg2013, Khatami2013, Beugeling2014, Steinigeweg2014, Sorg2014, Kim_strong2014, khodja_steinigeweg_15, beugeling_moessner_15, fratus_srednicki_15, chandaran_schulz_16, Mondaini2016, mondaini_rigol_17, lan_powell_17, garrison_grover_18, khaymovich_haque_19, hamazaki_ueda_19, jansen_stolpp_19, richter_gemmer_19, chan_deluca_19}.
The ETH is most commonly expressed by the Srednicki ansatz~\cite{srednicki_99} for the matrix elements of local observables $A_{nm}=\langle n|\hat A|m\rangle$ in the basis of eigenstates $\{ |n\rangle \}$ of the Hamiltonian $\hat H$.
The focus of this work are the diagonal matrix elements, which are (up to fluctuations) smooth functions of energies, $A_{nn}\simeq A(E_n)$.   An important feature beyond predictions of random matrix theory is that in general $A(E_n)\ne$ const.
In other words, $A_{nn}$ show some structure ~\cite{dalessio_kafri_16, hamazaki_ueda_18}, e.g., the slope of $A(E_n)$ is nonzero as shown in Fig.~\ref{fig0}. As a consequence of the structure, quantitative measures of fluctuations in finite systems, defined within energy windows that scale polynomially with system size, may be ambiguous. The most powerful indicator of the ETH established so far are ''local'' fluctuations of $A_{nn}$ (expressed in terms of nearest level fluctuations of $A_{nn}$), which decay exponentially with system size~\cite{Kim_strong2014, Mondaini2016, jansen_stolpp_19}.

In the context of quadratic and Bethe ansatz integrable models (shortly, integrable models), it is well known that fluctuations of most local observables decay slower than exponential, and hence the ETH is violated~\cite{pop1, Cassidy2011, vidmar16, rigol_16}.
This observation can be explained by arguing that Hamiltonian eigenstates with a similar distribution of local integrals of motion (LIOMs) also exhibit similar values of the diagonal matrix elements of observables.
Those arguments form the basis of the generalized ETH~\cite{Cassidy2011, He2013, vidmar16} and the quench action approach~\cite{caux_essler_13, wouters_denardis_14, brockmann_wouters_14, pozsgay_mestyan_14, mestyan_pozsgay_15, alba_calabrese_16, caux_2016}, which both provide a framework to explain the success of the generalized Gibbs ensemble to describe equilibration in  integrable systems~\cite{gge, vidmar16}.
In spite of those advances \cite{Takashi_2019}, nevertheless, the quantitative role of integrals of motion in the ETH remains widely unexplored.

In this Letter we develop a theory that quantifies the impact of integrals of motion on fluctuations and structure of the diagonal matrix elements of local observables (shortly, observables). 
First we show that a specific size--dependence of the latter fluctuations entails the presence of LIOMs.
Then, we outline a generic procedure, applicable to both integrable and nonintegrable models, in which observables are subtracted by their projections on LIOMs as well as products of LIOMs (that are few-body but nonlocal conserved operators).
This reduces fluctuations and/or the structure of the diagonal matrix elements, which is explicitly demonstrated in two examples.

{\it Preliminaries.}
We study translationally--invariant (TI)  chains with $L$ sites and discuss {\it traceless} TI observables, $\hat A = 1/\sqrt{L} \sum_{j=1}^L \hat a_j$, where $\hat a_j$ is the density of  
$\hat{A}$ supported in the vicinity of site $j$.
The choice of the prefactor $1/\sqrt{L} $ is uncommon but convenient since it yields the operators normalized, i.e., $ 0 < \lim_{L \rightarrow \infty} || \hat A || < \infty$.
The Hilbert-Schmidt norm $|| \hat A ||$ is defined as
\begin{equation}
|| \hat A ||^2=\langle \hat A \hat A \rangle = \frac{1}{Z}\sum_n   \langle n| \hat A^2 |n\rangle = \frac{1}{Z}\sum_{n,m}    |A_{nm}|^2 \, ,
\label{hsn}
\end{equation} 
where $Z$  is the Hilbert space dimension.
We introduce averaging over infinite time window, 
$\hat{\bar{A}} = \lim_{\tau \rightarrow \infty}\int_0^{\tau} {\rm d}t \hat A(t)/\tau$,
which projects out the off--diagonal matrix elements of the observable $\hat A$ such that $\hat{\bar{A}}=\sum_{n} A_{nn} |n\rangle \langle n|$.  
We define the operator stiffness $\sigma^2_A$ as the squared norm of the time-averaged operator 
\begin{equation} \label{def_spectrum_variance}
\sigma^2_A = \langle \hat{\bar A} \hat{\bar A} \rangle = \frac{1}{Z} \sum_n (A_{nn})^2 \,.
\end{equation}
Finally, we introduce fluctuations of diagonal matrix elements above their microcanonical average~\cite{Ikeda2013, Beugeling2014},
\begin{equation} \label{nETH}
\Sigma^{2}_A(\Delta)=\frac{1}{Z_\Delta} \sum_{E_n \in \Delta} \left[ A_{nn}- A (E_n)  \right]^2 \, ,
\end{equation}  
where $A(E_n)$ is the microcanonical average of $\hat A$ at energy $E_n$ and $Z_\Delta$ is the number of eigenstates in the energy window $ \Delta$.
If fluctuations are sampled over all eigenstates, then $Z_\Delta \to Z$ and $\Sigma^{2}_A(\Delta) \to \Sigma^{2}_A$.
The role of $A(E_n)$ is to remove the structure of the diagonal matrix elements, if present.
For observables with no structure [i.e., for $A(E_n) \to 0$], the stiffness $\sigma^2_A$ becomes identical to $\Sigma^{2}_A$.
An important property used throughout the paper is that $\Sigma^2_A$  increases if the microcanonical average $A(E_n)$ is replaced by other  smooth function $f(E_n)$,
\begin{equation} \label{inequality}
 \Sigma^2_A \; \le \;  \frac{1}{Z} \sum_{n} \left[ A_{nn}-  f(E_n) \right]^2  \, .
\end{equation}

Physically, we justify the normalization of observables and the use of stiffness recalling the results for ballistic particle transport in one-dimensional integrable models.
The ballistic transport shows up as  nonvanishing charge stiffness $\sigma^2_{I} > 0 $ in the thermodynamic limit~\cite{grabowski, zotos1996, zotos1997, zotos1999, zotos_unpub, benz,review2007, shastry, Sirker2009, tomaz_quasilocal11, Tomaz2011, prosen_1998, herbrych2011, Marko2011, my3, vidmar2013, robin2013, crivelli2014, mendoza2015}, defined  for the observable  $\hat I = 1/\sqrt{L} \sum_i \hat \jmath_i$, where $\hat \jmath_i$ is the charge current flowing between sites $i$ and $i+1$.
Since the microcanonical average $I(E_n)$ vanishes due to time--reversal symmetry, i.e., the observable $\hat I$ has no structure, the nonvanishing stiffness $\sigma^2_{I}$ also reflects nonvanishing fluctuations $\Sigma^2_I$ and the breakdown of the ETH~\cite{steinigeweg2013}.

We introduce a measure of violation of the ETH in integrable systems: 
fluctuations $\Sigma^2_A(\Delta)$ of a {\it normalized}, traceless TI observable $\hat A$ do not vanish, 
\begin{equation} \label{def_violation_eth}
 \lim_{L \to \infty} \Sigma^2_A(\Delta) > 0 \; ,
\end{equation} 
for arbitrary energy window $\Delta$.
We verify Eq.~(\ref{def_violation_eth}) for  other structureless operator in an interacting integrable model~\cite{suppmat}.
We also note that the validity of Eq.~(\ref{def_violation_eth}) is supported by results for  {\it intensive} observables $\hat A^{\rm int}= \hat A/\sqrt{L}$  in integrable models,  for which typically  $\Sigma^{2}_{A^{\rm int}} \sim 1/L$, as reported in Refs.~\cite{biroli2010, brandino_deluca_12, Ikeda2013, Alba2015}.
In this case, vanishing of  $\Sigma^{2}_{A^{\rm int}}$ can be viewed as a consequence of the vanishing operator norm,  $|| \hat A^{\rm int} ||^2 \sim 1/L$.
In fact, if $A(E_n) = 0$, Eqs.~(\ref{hsn}) and~(\ref{nETH}) imply $\Sigma^{2}_{A^{\rm int}} \leq || \hat A^{\rm int} ||^2$.

{\it Violation of ETH entails existence of LIOMs.}
Violation of the ETH, as defined in Eq.~(\ref{def_violation_eth}), implies [together with Eq.~(\ref{inequality})] an inequality
\begin{equation} \label{nETH1}
 0 \; < \; \frac{1}{Z} \sum_{n} \left[ A_{nn}-  f(E_n) \right]^2  \, ,
\end{equation}   
which in the thermodynamic limit holds for arbitrary smooth function of energy $f(E_n)$.
We show in what follows that Eq.~(\ref{nETH1})  entails the presence of LIOMs.
We introduce a {\it projected} observable
\begin{equation} \label{aperp}
\hat A_{\perp} = \hat A - p_A \hat H \,, \quad \quad p_A=\frac{ \langle \hat A \hat H \rangle}{\langle \hat H \hat H \rangle} \, ,
\end{equation} 
and argue that the time-averaged observable $\hat{\bar{A}}_{\perp}$ is a LIOM, orthogonal to the Hamiltonian $\hat H$.

We first recall that any time-averaged operator is conserved (but not necessarily local), and that time averaging is an orthogonal projection
$\langle \hat{\bar{A}} \hat B \rangle = \langle \hat A \hat{\bar{B}} \rangle =  \langle \hat{\bar{A}} \hat{\bar{B}} \rangle$, where $\langle \hat A \hat B\rangle$ is the Hilbert-Schmidt scalar product of operators $\hat A$ and $\hat B$,
see Eq. (\ref{hsn}).
Then, orthogonality of $\hat{\bar{A}}_{\perp}$ to $\hat H$ follows from orthogonality of $\hat A_{\perp}$ to $\hat H$ in construction of Eq.~(\ref{aperp}), since
$\langle \hat H \hat{\bar{A}}_{\perp}  \rangle = \langle \hat{\bar{H}} \hat{A}_{\perp}  \rangle = \langle \hat{H}  \hat{A}_{\perp} \rangle = 0$.

The key step is to show locality of $\hat{\bar{A}}_{\perp}$.
In general, testing locality of integrals of motion via analyzing their supports is  a tough problem.
However, it is known that in the thermodynamic limit only LIOMs (including pseudolocal conserved operators~\cite{tomaz_quasilocal11}) contribute to the Mazur bound~\cite{mazur,zotos1997} 
\begin{equation} \label{mazur}
\langle  \hat{\bar{A}} \hat{\bar{A}} \rangle \ge \sum_{\alpha} \frac{\langle \hat A \hat Q_{\alpha} \rangle^2}{\langle \hat Q_{\alpha} \hat Q_{\alpha} \rangle } \, .
\end{equation}   
Hence any conserved operator $\hat Q_{\alpha}$ is local (or pseudolocal) when 
$\langle \hat A \hat Q_{\alpha} \rangle^2/ \langle \hat Q_{\alpha} \hat Q_{\alpha} \rangle >0 $ for some normalized, TI and local  observable $\hat A$.
The latter concept relaxes the constraint on strictly local densities of LIOMs and is equivalent to the definition of pseudolocality  introduced in Ref.~\cite{Ilievski_2016}.
Using the identity
$\langle \hat{\bar{A}}_{\perp} \hat{\bar{A}}_{\perp} \rangle= \langle \hat A_{\perp} \hat{\bar{A}}_{\perp} \rangle= \langle (\hat A - p_A \hat H) \hat{\bar{A}}_{\perp} \rangle = \langle \hat A \hat{\bar{A}}_{\perp} \rangle$
we get
\begin{equation} \label{norm3}
\frac{ \langle \hat A \hat{\bar{A}}_{\perp} \rangle^2 }{\langle \hat{\bar{A}}_{\perp} \hat{\bar{A}}_{\perp} \rangle} = \langle \hat{\bar{A}}_{\perp} \hat{\bar{A}}_{\perp} \rangle = \sigma_{A_\perp}^2 \, .
\end{equation}
Inequality~(\ref{nETH1}) implies that the stiffness of $\hat A_\perp$ is nonzero,
\begin{equation}
\sigma_{A_\perp}^2 = \langle \hat{\bar{A}}_{\perp} \hat{\bar{A}}_{\perp} \rangle=  \frac{1}{Z} \sum_n \left( A_{nn}-p_A E_n \right)^2 >0,
\end{equation}
hence the conserved operator $\hat{\bar{A}}_{\perp}$ is local or pseudolocal.

{\it Testing completeness of a set of LIOMs.} 
We next consider a system which contains an orthogonal set of  LIOMs $\{ \hat Q_{\alpha} \}$, $\langle \hat Q_{\alpha} \hat Q_{\beta} \rangle \propto \delta_{\alpha \beta}$, and we generalize the definition of projected observables, introduced in Eq.~(\ref{aperp}), to
\begin{equation} \label{perdef}
\hat A_{\perp} = \hat A - \sum_{\alpha} p_{A\alpha} \hat Q_{\alpha} \,, \quad \quad p_{A\alpha} = \frac{ \langle \hat A \hat Q_\alpha \rangle}{\langle \hat Q_\alpha \hat Q_\alpha \rangle} \, .
\end{equation} 
The observables $\hat A_{\perp}$ and $\hat{\bar{A}}_{\perp}$ are orthogonal to all LIOMs $\{ \hat Q_\alpha \}$ since
$\langle \hat{\bar{A}}_{\perp} \hat Q_{\beta} \rangle = \langle \hat A_{\perp} \hat Q_{\beta} \rangle = \langle \hat A \hat Q_{\beta} \rangle - p_{A\beta}  \langle  \hat Q_{\beta} \hat Q_{\beta} \rangle = 0$, and as a consequence, Eq.~(\ref{norm3}) is still valid.  If the set of LIOMs is complete, then the norm of  $\hat{\bar{A}}_\perp$  must vanish in the thermodynamic limit, and hence
\begin{equation}   \label{def_sigma_Aperp}
 \lim_{L\to \infty} \sigma_{A_\perp}^2 = 0. 
 \end{equation}
Otherwise, $\hat{\bar{A}}_{\perp}$ is local (or pseudolocal) and it represents an additional LIOM which is missing in the set $\{ \hat Q_{\alpha} \}$. 

{\it Size-dependence of the stiffness.}
As an important outcome of previous analysis we introduced projected operators $\hat A_{\perp}$, which are still local operators but their stiffnesses vanish in the thermodynamic limit.
They can be constructed for both integrable and nonintegrable models.
For the latter, $\hat H$ is the only LIOM and the stiffness vanishes for $\hat A_{\perp}$ defined in Eq.~(\ref{aperp}).

We conjecture that the stiffness may be further reduced if operators are additionally subtracted by their projections on products of LIOMs, which are few-body nonlocal operators.
Below we provide analytical and numerical evidence for our conjecture.

\begin{figure}
\centering
\includegraphics[width=\columnwidth]{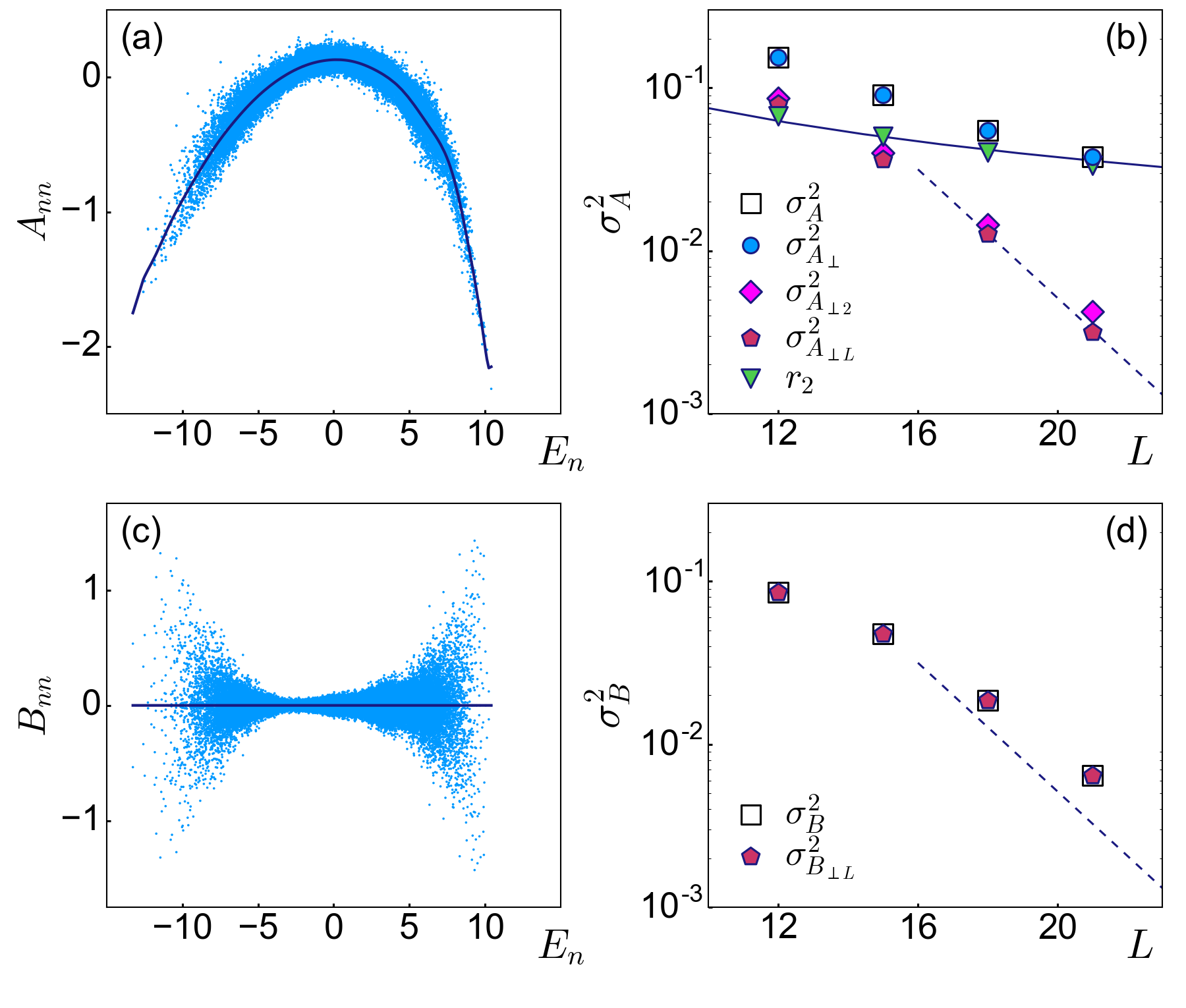}
\caption{
Diagonal matrix elements of observables $\hat A$ and $\hat B$ (see text for definitions) for the generic Hamiltonian $\hat H$~(\ref{ham1}).
Symbols in (a) and (c) show $A_{nn}$ and $B_{nn}$, respectively, versus $E_n$ for $L=21$ sites, while lines represent $f_{k=L}(E_n)$~(\ref{Ak_generic}).
(b) and (d): stiffness $\sigma^2_A$ and $\sigma^2_B$, respectively, and the corresponding $k$--stiffness $\sigma^2_{A_{\perp k}}$ and $\sigma^2_{B_{\perp k}}$, where $\sigma^2_{A_{\perp 1}} \equiv \sigma^2_{A_{\perp}}$.
In (b) we also plot the projection $r_2$~(\ref{noise}).
Solid and dashed lines are guides to the eye, given by polynomial ($\sim 1/L$) and exponential ($\sim e^{-L/2.2}$) functions, respectively.
}
\label{fig1}
\end{figure} 

{\it Products of LIOMs in generic systems.}
We first study a generic system where the only LIOM is the Hamiltonian $\hat H$, and hence the only products of LIOMs are the powers of $\hat H$.
We start by finding the polynomial $f(E_n)$ that minimizes the right hand side of Eq.~(\ref{inequality}).
This is the best polynomial fit to the microcanonical average $A(E_n)$. 
We introduce a polynomial of degree $k$ of the Hamiltonian (also denoted as the $k$-product of Hamiltonian),
\begin{equation} \label{H_kproduct}
\hat H_{\perp k} = \hat H^k-\sum_{j=1}^{k-1} \frac{\langle \hat H^k \hat H_{\perp j}\rangle}{\langle \hat H_{\perp j} \hat H_{\perp j}\rangle } \hat H_{\perp j},
\end{equation}
where $\hat H_{\perp 1} = \hat H$.
The $k$-products are orthogonal by construction, $\langle \hat H_{\perp k} \hat H_{\perp l}\rangle \propto \delta_{k,l}$.
Then, the central step is to construct $k$-projected observables $\hat A_{\perp k}$,
\begin{equation} \label{Ak_generic}
\hat A_{\perp k} = \hat A - \hat f_k(\hat H) \, , \quad \hat f_k( \hat H) =  \sum_{j=1}^k  \frac{\langle \hat A \hat H_{\perp j}\rangle}{\langle \hat H_{\perp j} \hat H_{\perp j} \rangle} \hat H_{\perp j} \, ,
\end{equation}
which can be seen as a generalized form of Eq.~(\ref{perdef}), with $\hat A_{\perp 1} \equiv \hat A_{\perp}$.
Using orthogonality of $\hat H_{\perp k}$ one easily finds an explicit form of the $k$--stiffness,
\begin{equation} \label{noise}
\sigma^2_{A_{\perp k}} = \sigma_A^2 - \sum_{j=1}^k r_j \, , \quad \quad  r_j = \frac{\langle \hat A \hat H_{\perp j}\rangle^2}{\langle \hat H_{\perp j} \hat H_{\perp j} \rangle} \, , 
\end{equation}
where the stiffness at $k=1$ is $\sigma^2_{A_{\perp 1}} \equiv \sigma^2_{A_\perp}$. 
In~\cite{suppmat} we show that  $f_k(E_n) = \langle n | \hat f_k(\hat H) |n\rangle$   is indeed the best polynomial fit to $A(E_n)$ for a given degree $k$, hence  $\lim_{k \rightarrow \infty} f_{k} (E_n)=A(E_n)$. 

It follows from Eq.~(\ref{noise})  that the stiffness $\sigma_A^2$ is bounded from below by all  $r_k$, i.e., by  the projections of $\hat A$ on $k$--th power of the Hamiltonian.
Using a Gaussian density of states one can show~\cite{suppmat} that  $r_k \sim {\cal O}(1/L^{k-1})$.
Below, we demonstrate that the stiffness $ \sigma_A^2$ may be reduced order by order via subtracting these projections, i.e., via considering operators $\hat A_{\perp k}$ introduced in Eq.~(\ref{Ak_generic}), for which the leading term of the $k$--stiffness $\sigma^2_{A_{\perp k}}$ is at most of the order ${\cal O}(1/L^{k})$.  The physical picture behind our construction is that the diagonal matrix elements of $k$-projected observables $\hat A_{\perp k}$ become structureless, i.e., they become closer to the ones typical for the random matrix theory.

For this sake, we study a  nonintegrable periodic chain of interacting spinless fermions on $L$ sites and with $N=L/3$ particles,
\begin{equation}
\hat H = -\sum_{j=1}^L (e^{i \phi} \hat c^{\dagger}_{j+1} \hat c_j + {\rm H.c.})
+ \sum_{j=1}^L (V \hat{\tilde{n}}_j \hat{\tilde{n}}_{j+1} + W \hat{\tilde{n}}_j \hat{\tilde{n}}_{j+2}) \, .
\label{ham1}
\end{equation}
Here, $\hat n_j= \hat c^{\dagger}_{j} \hat c_j$,  $\hat{\tilde n}_j = \hat n_j -1/3$ and we set $V=W=1$. 
The Hamiltonian (\ref{ham1}) has been diagonalized separately in  each sector with total momentum $q$.
We remove degeneracies in all $q$--sectors by introducing a flux $\phi=2\pi/L$ and $\pi/L$ for even and odd $N$, respectively.

We study two observables: the generalized hopping energy
$\hat A = \frac{1}{\sqrt{L}} \sum_{j}  (\hat{\kappa}_j + \hat{\kappa}^{\dagger}_j) $,
and the generalized current
$\hat B = \frac{1}{\sqrt{L}} \sum_{j}  i(\hat{\kappa}_j - \hat{\kappa}^{\dagger}_j)$, with $\hat{\kappa}_j=e^{2 i \phi} \hat c^{\dagger}_{j+1}(1-3 \hat n_j) \hat c_{j-1}$.
In Figs.~\ref{fig1}(a) and~\ref{fig1}(c) we show their diagonal matrix elements $A_{nn}$ and $B_{nn}$, respectively, and the corresponding polynomial fits $f_{k=L}(E_n)$ from Eq.~(\ref{Ak_generic}).

A common feature of both observables is that they have zero projection on $\hat H$ [i.e., $r_1 = 0$ in Eq.~(\ref{noise})], and hence $\sigma^2_A = \sigma^2_{A_\perp}$ and $\sigma^2_B = \sigma^2_{B_\perp}$.
In contrast, the observable $\hat A$ has nonzero while the observable $\hat B$ has zero projection on the 2-product of $\hat H$.
This explains the origin of the power-law decay ($\propto 1/L$) of $\sigma^2_{A_\perp}$ in Fig.~\ref{fig1}(b) (see also results for $r_2$, which approach $\sigma^2_{A_\perp}$ for large $L$).
After the contribution from $r_2$ is subtracted from the operator, we observe a nearly exponential decay of $\sigma^2_{A_{\perp k}}$ with $L$ for $k\geq 2$ in Fig.~\ref{fig1}(b). 
We note also that $\hat{B}$ has no projection on any power of the Hamiltonian, i.e., the diagonal matrix elements have no structure, and  $\sigma^2_{B}$ decays exponentially with $L$.

\begin{figure}
\centering
\includegraphics[width=\columnwidth]{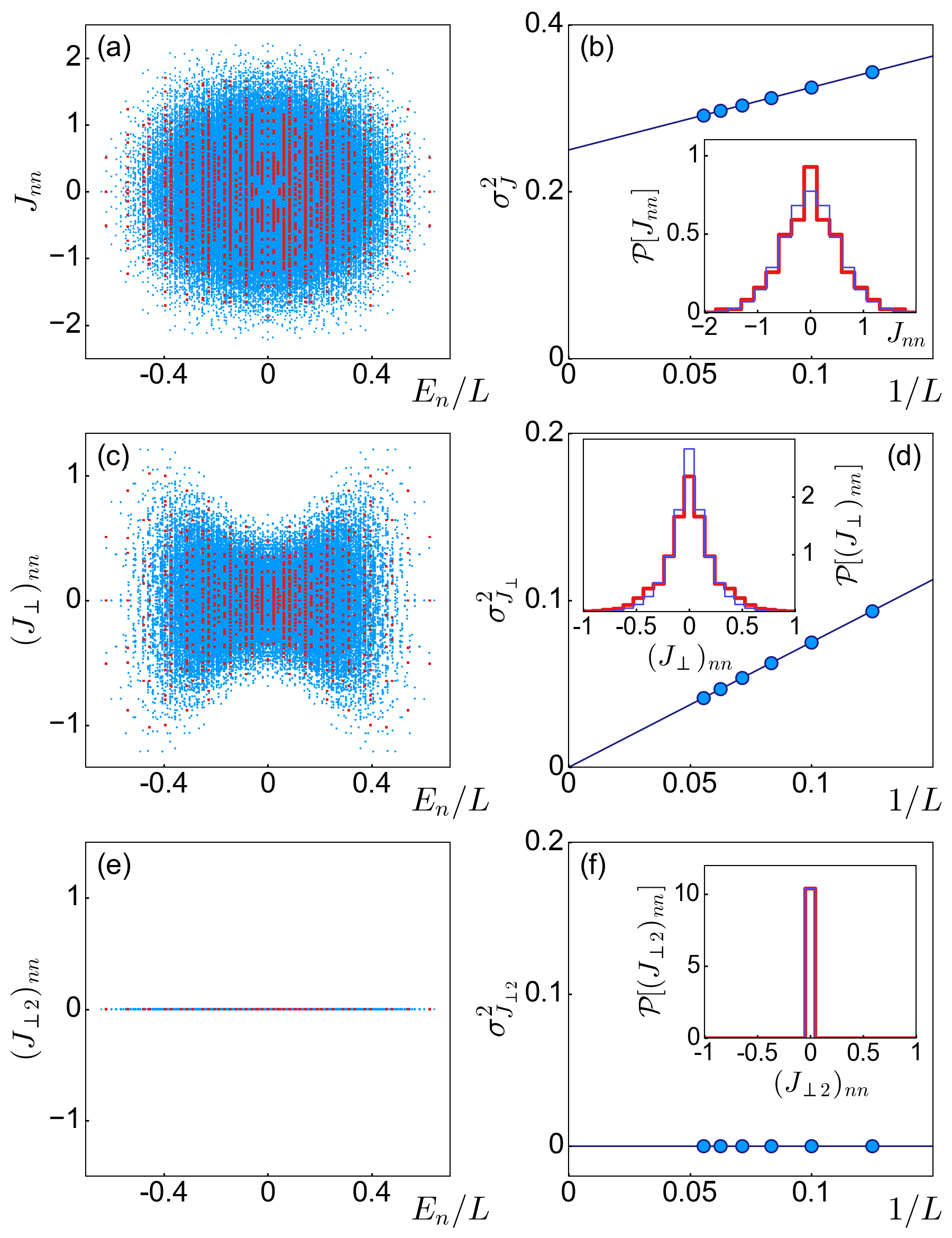}
\caption{Diagonal matrix elements of the observable $\hat J$, projected observable $\hat J_\perp$, and 2-projected observable $\hat J_{\perp 2}$, for the integrable HCBs Hamiltonian $\hat H_{\rm HCB}$ that includes all particle sectors (see text for details).
(a), (c) and (e): $J_{\rm nn}$, $(J_\perp)_{nn}$ and $(J_{\perp 2})_{nn}$, respectively, for $L=18$ (bright blue symbols) and $L=12$ (dark red symbols).
(b), (d) and (f): stiffness $\sigma^2_{J}$ and $k$-stiffness $\sigma^2_{J_\perp} \equiv \sigma^2_{J_{\perp 1}}$ (for $k=1$) and $\sigma^2_{J_{\perp 2}}$ (for $k=2$), respectively.
The insets show histograms of the corresponding distributions of the matrix elements.
Lines in (b) and (d) are the functions $1/4 + (3/4) L^{-1}$ and $(3/4) L^{-1}$, respectively.
}   
\label{fig2}
\end{figure} 

{\it Products of LIOMs in integrable systems.}
We now turn our focus to integrable models, for which the set of products of LIOMs is much richer.
A general $k$-product of LIOMs should be built iteratively, in analogy to Eq.~(\ref{H_kproduct}).
Note that the entire set of products of LIOMs should be orthogonal.
In the case of 2-products of LIOMs, denoted by $\hat X_{\gamma(\alpha,\beta)} = \hat Q_{\alpha} \hat Q_{\beta} - \langle \hat Q_{\alpha} \hat Q_{\beta} \rangle $, this is achieved by
\begin{equation} \label{def_2products_LIOMs}
\hat X_{\perp \gamma} = \hat X_{\gamma} - \sum_{\alpha}  \frac{\langle \hat X_{\gamma} \hat Q_{\alpha}  \rangle}{ \langle \hat Q_{\alpha} \hat Q_{\alpha} \rangle } \hat Q_{\alpha} - \sum_{\gamma'=1}^{\gamma-1}  \frac{\langle \hat X_{\perp \gamma} \hat X_{\perp \gamma'}  \rangle}{\langle \hat X_{\perp \gamma'} \hat X_{\perp \gamma'} \rangle } \hat X_{\perp \gamma'} \, .
\end{equation}

As an example, we study a chain of hard-core bosons (HCBs) with the Hamiltonian $\hat H_{\rm HCB} = - \sum_j (\hat b_{j+1}^\dagger \hat b_j + {\rm h.c.})$ using periodic boundaries and the onsite constraints $(\hat b_j^\dagger)^2 = (\hat b_j)^2 = 0$, where $\hat b_j^\dagger$ ($\hat b_j$) creates (annihilates) a boson on site $j$.
A complete set of LIOMs $\{ \hat Q_\alpha \}$ is given by noninteracting spinless fermions onto which the HCBs are mapped (see~\cite{suppmat} for details).

We construct a two-body  structureless observable
$\hat J = \sqrt{\frac{2}{L}} \sum_j \left( i \hat b_{j+1}^\dagger \hat n_j \hat b_{j-1} + {\rm H.c.} \right)$,
for which the microcanonical average vanishes, $J(E_n)=0$.
Figure~\ref{fig2}(a) shows the diagonal matrix elements $J_{nn}$ for two system sizes $L$, and Fig.~\ref{fig2}(b) shows that the stiffness $\sigma^2_J$ extrapolates to a nonzero value in the thermodynamic limit $L \to \infty$.
These results signal violation of the ETH as stated in Eq.~(\ref{def_violation_eth}) and the existence of LIOMs.

We then construct a projected observable $\hat J_\perp$ according to Eq.~(\ref{perdef}) using a complete set of LIOMs.
Figure~\ref{fig2}(c) reveals that the support of the diagonal matrix elements of $(J_\perp)_{nn}$ is reduced when compared to $J_{nn}$ in Fig.~\ref{fig2}(a).
Moreover, Fig.~\ref{fig2}(d) shows a vanishing stiffness $\sigma^2_{J_\perp} \propto 1/L$, in agreement with Eq.~(\ref{def_sigma_Aperp}).

Finally, we construct the 2-projected observable $\hat J_{\perp 2}$, which is a generalization of Eq.~(\ref{Ak_generic}) for $k=2$ to include all the possible 2-products of LIOMs $\hat X_{\perp \gamma}$ from Eq.~(\ref{def_2products_LIOMs}).
Remarkably, all the diagonal matrix elements of $(J_{\perp 2})_{nn}$ are exactly zero already in finite systems, as shown in Figs.~\ref{fig2}(e) and~\ref{fig2}(f).
This reveals a special instance of the ETH, where the diagonal matrix elements form a well-defined function with zero fluctuations at any system size.

{\it Conclusions.} 
In this Letter we made steps towards a unified treatment of the ETH in integrable and generic quantum systems.
We introduced a protocol to construct projected observables, i.e., observables subtracted by their projections on LIOMs and products of LIOMs. In  the case of  generic nonintregrable systems, these observables gradually become  
structureless (albeit nonlocal)  similarly to the quantities in the random matrix theory. We have demonstrated that fluctuations of the diagonal matrix elements of projected observables decay exponentially with the system size. 
Finally, we conjecture for integrable systems that this approach  eliminates not only the structure but also fluctuations of the diagonal matrix elements which originate from projections on products of LIOMs others than  Hamiltonian. So far our applications concerned translationally invariant systems, and extensions to models without translational invariance, e.g. disordered systems, are desired for future work.

\acknowledgements
We acknowledge discussions with A. Polkovnikov, T. Prosen and M. Rigol.
This work is supported by the National Science Centre, Poland via Projects 2016/23/B/ST3/00647 (M.M.), and the Slovenian Research Agency (ARRS), Research core fundings No.~P1-0044 and No.~J1-1696 (L.V.).

\bibliographystyle{biblev1}
\bibliography{references}
 
\newpage
\phantom{a}
\newpage
\setcounter{figure}{0}
\setcounter{equation}{0}

\renewcommand{\thetable}{S\arabic{table}}
\renewcommand{\thefigure}{S\arabic{figure}}
\renewcommand{\theequation}{S\arabic{equation}}

\renewcommand{\thesection}{S\arabic{section}}

\onecolumngrid

\begin{center}

{\large \bf Supplemental Material:\\
Eigenstate thermalization hypothesis and integrals of motion}\\

\vspace{0.3cm}

Marcin Mierzejewski$^{1}$ and Lev Vidmar$^{2}$\\
$^1${\it Department of Theoretical Physics, Faculty of Fundamental Problems of Technology, \\ Wroc\l aw University of Science and Technology, 50-370 Wroc\l aw, Poland}\\
$^2${\it Department of Theoretical Physics, J. Stefan Institute, SI-1000 Ljubljana, Slovenia}

\end{center}

\vspace{0.6cm}

\twocolumngrid

\label{pagesupp}

\section{Observables in integrable systems} \label{app1}

Here we study the scaling of fluctuations and stiffnesses of the diagonal matrix elements of operators in an integrable chain of interacting spinless fermions, and discuss normalization of operators.
We consider the Hamiltonian in Eq.~(\ref{ham1}) (see the main text) with $W=0$, diagonalized separately in each sector with total momentum $q$.
All other parameters, including the particle filling $N/L=1/3$, remain the same as in the main text.
We study two normalized current operators (i.e., operators that have a $L$-independent norm),
\begin{eqnarray}
\hat{o}_j &=&i e^{2 i \phi} \hat c^{\dagger}_{j+1} \hat c_{j-1}  + {\rm H.c.}  \label{Adef} \,, \\
\hat{O} &= & \frac{\sqrt{L-1}}{L} \sum_{j} \hat{o}_j \label{Bdef} \,,
\end{eqnarray}
for which microcanonical averages vanish, i.e., $o_j(E_n) = O(E_n)=0$.

\begin{figure}
\centering
\includegraphics[width=\columnwidth]{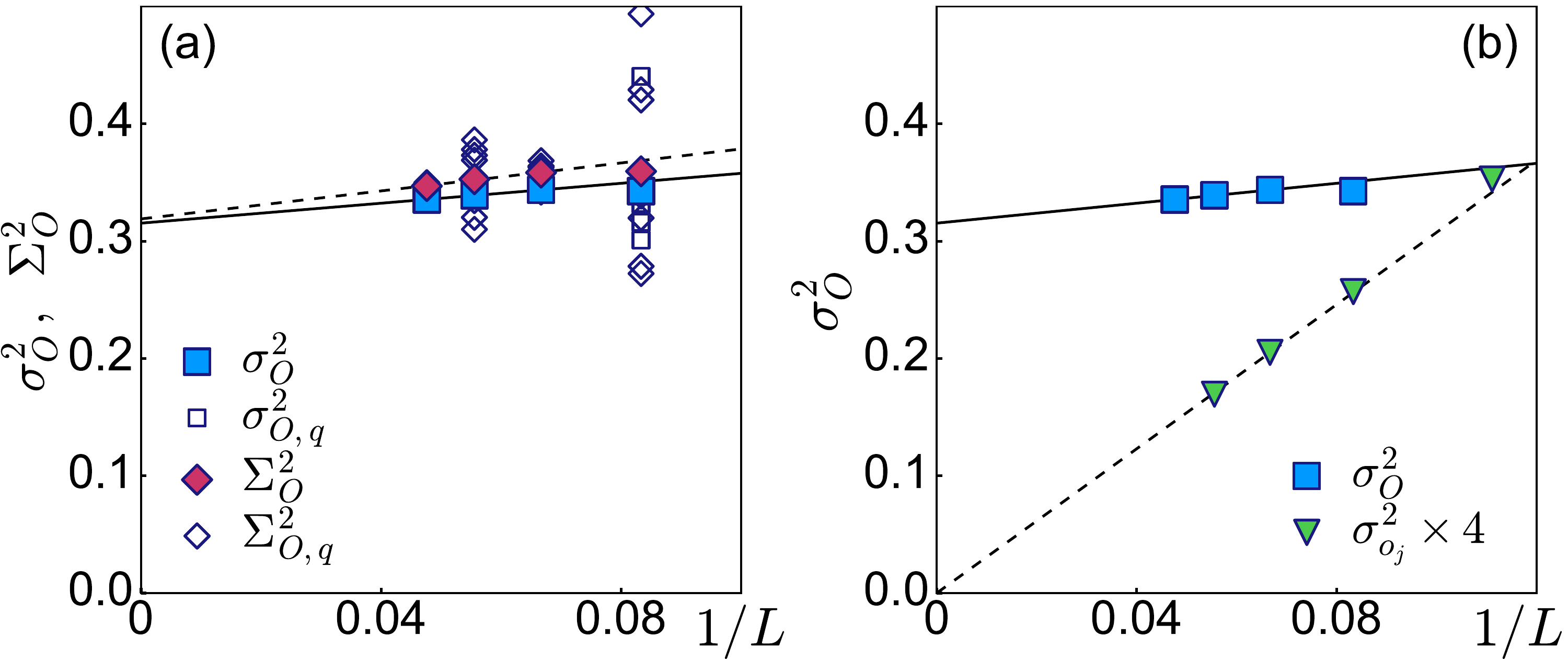}
\caption{
(a) Filled symbols represent the stiffness $\sigma^2_O$ and the fluctuations $\Sigma^2_O(\Delta)$ of the observable $\hat O$ from Eq.~(\ref{Bdef}).
Open symbols represent the same quantity, but calculated in a single $q$-sector only.
(b) Stiffness $\sigma^2_O$ and $\sigma^2_{o_j}$ of observables $\hat O$~(\ref{Bdef}) and $\hat o_j$~(\ref{Adef}), respectively.
Solid and dashed lines are fits to the function $\alpha + \beta/L$ [we set $\alpha=0$ for the dashed line in (b)].
All fits of the data correspond to three largest $L$ points.
}
\label{figS1}
\end{figure}

Figure~\ref{figS1}(a) [filled symbols] shows the fluctuations $\Sigma^2_O(\Delta)$, see Eq.~(\ref{nETH}), and the stiffness $\sigma^2_O$, see Eq.~(\ref{def_spectrum_variance}), for the observable $\hat O$.
The fluctuations are calculated in the energy interval $\Delta \in [E_\infty - \delta, E_\infty + \delta]$ using $\delta =3$, where $E_{\infty}$ is the energy that corresponds to infinite temperature.
Open symbols in Fig.~\ref{figS1}(a) show results in a single $q$-sector, which we denote by $\Sigma^2_{O,q}$ and $\sigma^2_{O,q}$.
Results confirm that these quantities do not vanish in the thermodynamic limit and are consistent with Eq.~(\ref{def_violation_eth}) in the main text.

In Fig.~\ref{figS1}(b) we compare the stiffnesses $\sigma^2_O$ and $\sigma^2_{o_j}$ of observables $\hat O$ and $\hat o_j$, respectively.
While $\sigma^2_O$ remains finite in the thermodynamic limit, $\sigma^2_{o_j}$ decays as $1/L$, despite both operators being normalized.
The essential difference between these operators is that only $\hat{O}$ is translationally invariant.
In this sense, $\hat{O}$ is compatible with the symmetries of the Hamiltonian and with the translationally invariant LIOMs.
The operator $\hat{o}_j$ is, in contrast, not translationally invariant.
For each translationally invariant LIOM  $\hat{Q}_{\alpha} = \sum_i  \hat{q}_{\alpha,i}$ one finds the projection
\begin{equation}
\frac{\langle \hat{o}_j \hat{Q}_{\alpha} \rangle^2}{ \langle \hat{Q}_{\alpha} \hat{Q}_{\alpha} \rangle}
\sim  \frac{  \langle \hat{o}_j \hat{q}_{\alpha,j} \rangle^2 }{\sum_{i}  \langle \hat{q}_{\alpha,i} \hat{q}_{\alpha,i}  \rangle } \, .
\end{equation}
This projection is at most of the order $O(1/L)$ and hence none of the translationally invariant LIOM can be derived from the stiffness of $\hat{o}_j$. 

To summarize, we note that if the stiffness of a normalized operator does not vanish in the thermodynamic limit, then it entails the presence of a LIOM.
However, the opposite claim does not hold true.
When an operator is incompatible with the structure of LIOMs, e.g. with respect to translational symmetry, its stiffness may vanish (as a power law) in the thermodynamic limit despite the presence of LIOMs.
 
\section{Products of LIOMs in generic systems} \label{app2}

{\it Best polynomial fit to the microcanonical average.}
The  best fit  to  $A(E_n)$ via the polynomial of degree $k$ is given by the matrix elements
$f_k(E_n)=\langle n | \hat{f}_k(\hat H) |n \rangle $, where
\begin{equation}
\hat{f}_k(\hat H) = \sum_{j=1}^k  \frac{\langle \hat{A} \hat{H}_{\perp j}\rangle}{\langle \hat{H}_{\perp j} \hat{H}_{\perp j} \rangle} \hat{H}_{\perp j}, \label{fit}
\end{equation}
and the $j$-products of $\hat{H}$, denoted as $\hat{H}_{\perp j}$, are defined by Eq.~(\ref{H_kproduct}) in the main text.
In order show that Eq.~(\ref{fit}) is indeed the best fit, one needs to show that this particular function $f$ minimizes the stiffness of $\hat{A}_{\perp k}=\hat{A}-\hat{f}_k( \hat H)$,
\begin{equation}
\sigma^2_{A_{\perp k}}= \langle \hat{\bar{A}}_{\perp k} \hat{\bar{A}}_{\perp k} \rangle =
\frac{1}{Z} \sum_{n} \left[ A_{nn}-  f_k(E_n) \right]^2  \, .
\end{equation}

We note that  $\hat{A}_{\perp k }$ is orthogonal to  $\hat{H}_{\perp i}$ for all  $i \le k$, hence  
$\langle  \hat{\bar{A}}_{\perp k}  \hat{H}_{\perp i} \rangle=\langle  \hat{A}_{\perp k}  \hat{H}_{\perp i} \rangle=0$. 
Moreover, all $ \hat{H}_{\perp i}$ are mutually orthogonal by construction,   $\langle   \hat{H}_{\perp j}   \hat{H}_{\perp i} \rangle \propto \delta_{ij}$.
One may modify coefficients of the polynomial, $\hat{f}'_k(H) = \hat{f}_k(H)+ \sum_j \alpha_j \hat{H}_{\perp j}$ and introduce the corresponding projected operator $\hat{A}'_{\perp k}=\hat{A}-\hat{f}'_k(H)$.
Then,
\begin{eqnarray}
&& \frac{1}{Z} \sum_{n} \left[ A_{nn}-  f'_k(E_n) \right]^2= \langle \hat{\bar{A}}'_{\perp k} \hat{\bar{A}}'_{\perp k} \rangle   \nonumber \\
&& =\langle (\hat{\bar{A}}_{\perp k}- \sum_j \alpha_j \hat{H}_{\perp j})^2 \rangle 
= \sigma^2_{A_{\perp k}}+ \sum_j \alpha^2_j  || \hat{H}_{\perp j} ||^2.   \nonumber \\
\end{eqnarray}
Therefore, any modification of the polynomial in Eq.~(\ref{fit})  increases the stiffness of the projected operator.

{\it Projections on powers of Hamiltonian.}
Here, we discuss the $L$--dependence of the projection $r_i$, as defined via Eq.~(\ref{noise}) in the main text, for a generic nonintegrable system. Our main assumption is that the many-body density of states is a Gaussian function
 \begin{equation}
 \rho(E)=\frac{1}{Z} \sum_n \delta(E-E_n)=\frac{\exp \left( \frac{-E^2}{2 L \varepsilon_0^2} \right)}{\sqrt{2 \pi L  \varepsilon_0^2}},
 \end{equation}  
where, for simplicity, we put $ \varepsilon_0=1$. We consider diagonal matrix elements of the $k$-products of $\hat{H}$, as defined via Eq.~(\ref{H_kproduct}) in the main text. 
Since, $\langle  \hat{H}_{\perp i} \hat{H}_{\perp j}\rangle \propto \delta_{ij}$, it is useful to define  normalized operators as
 \begin{equation}
 \tilde{H}_{\perp i}= \frac{\hat{H}_{\perp i}}{\sqrt{\langle \hat{H}_{\perp i} \hat{H}_{\perp i} \rangle }} \,\,,\quad \quad \langle  \tilde{H}_{\perp i} \tilde{H}_{\perp j}\rangle = \delta_{ij} \,. \label{ort}
 \end{equation} 
As it follows from  Eq.~(\ref{H_kproduct}) in the main text, $\tilde{H}_{\perp i}=\hat{w}_i(\hat H)$, where $w_i$ is a polynomial of degree $i$. Using the orthogonality relation, Eq.~(\ref{ort}), together with the density of states
we find that polynomials $w_i$ are orthogonal with the Gaussian weight function
\begin{equation} 
 \int_{-\infty}^{\infty} {\rm d} E \; w_i(E) w_j(E) \frac{\exp \left( \frac{-E^2}{2 L} \right)}{\sqrt{2 \pi L}} = \delta_{ij}.
\end{equation}
Comparing the latter equation with the orthogonality relation for the Hermite polynomials ${\cal H}_i$, 
 \begin{equation}
  \int_{-\infty}^{\infty} {\rm d} x  {\cal H}_i(x)  {\cal H}_j(x) \exp(-x^2)= \delta_{ij} \sqrt{\pi} (i !) 2^i,
 \end{equation} 
 one finds for the Gaussian density of state that  $\tilde{H}_{\perp i}$ is the Hemite polynomial of $\hat{H}$ of degree $i$
\begin{equation}
w_i(\hat H)=\frac{ {\cal  H}_i \left(\frac{\hat H}{\sqrt{2L}} \right) }{\sqrt{2^i i !}}.
\end{equation} 
 
We  consider a translationally invariant normalized observable $\hat{A}$ that fulfills the ETH. Namely, we assume that  its diagonal matrix elements are determined by the energy density
up to exponentially decaying fluctuations,
 \begin{equation}
 \langle n | \hat{A} |n \rangle=\sqrt{L} \; {\cal A}\left(\frac{E_n}{L} \right) +O_n(e^{-L}),
 \end{equation}
where the term $\sqrt{L}$ originates from normalization discussed in the main text. In order to obtain the projection $r_i$  defined in the main text in Eq.~(\ref{noise}) we calculate
\begin{eqnarray}
\tilde{r}_i &= & \sqrt{r_i}=\langle A \tilde{H}_{\perp i} \rangle \\  
&=   &  \int_{-\infty}^{\infty} {\rm d} E \;  \sqrt{L} \; {\cal A}\left(\frac{E}{L} \right)  w_i(E)  \frac{\exp \left( \frac{-E^2}{2 L} \right)}{\sqrt{2 \pi L}}, \nonumber \\
& =  & \sqrt{\frac{L}{\pi}} \int_{-\infty}^{\infty} {\rm d} x  {\cal A}\left( x \sqrt{2/L} \right) \frac{ {\cal H}_i(x)}{\sqrt{2^i i !}}\exp(-x^2) \, . \nonumber
\end{eqnarray}
We expand the microcanonical average  ${\cal A}\left( x \sqrt{2/L} \right)$ in power series. We also use the fact that the Hermite polynomial  ${\cal H}_i(x)$ is orthogonal to all polynomials
of degree $j<i$.  Therefore, the leading contribution to $\tilde{r}_i$ comes from the $i$-th order expansion of ${\cal A}\left( x \sqrt{2/L} \right)$ and, therefore,  scales with $L$ according to
\begin{equation}
\tilde{r}_i \sim \left(\frac{1}{\sqrt{L}}\right)^{i-1}.
\end{equation}
Finally, one finds that the projections which determine the stiffness  decay with $L$ according to the power-law
\begin{equation}
r_i = \tilde{r}^2_i  \sim  \frac{1}{L^{i-1}}.
\end{equation}

\section{LIOMs in integrable systems} \label{app3}

In the main text, we also study integrable HCBs, which can be mapped onto spins 1/2 and spin 1/2 onto spinless fermions as $\hat b_j = e^{i \pi \sum_{m < j} \hat c_m^\dagger \hat c_m}$, such that the Hamiltonian $\hat H_{\rm HCB}$ maps onto a Hamiltonian for noninteracting spinless fermions, $\hat H_{\rm SF} = - \sum_j (\hat c_{j+1}^\dagger \hat c_j + {\rm h.c.})$.
For the latter, we define a set of orthogonal and normalized conserved one-body operators as
\begin{eqnarray}
 &\hat T^{(0)} & =  \frac{1}{\sqrt{L}} \left( 2\hat N - L \right) \,, \\
 &\hat T^{(n)} &=  - \sqrt{\frac{2}{L}} \sum_j \left( \hat c_j^\dagger \hat c_{j+n} + \hat c_{j+n}^\dagger \hat c_j \right) \,, \\
 &\hat J^{(n)}    & =  - i  \sqrt{\frac{2}{L}} \sum_j \left( \hat c_j^\dagger \hat c_{j+n} - \hat c_{j+n}^\dagger \hat c_j \right) \,, \\
 &\hat T^{(L/2)} & =  -  \frac{1}{\sqrt{L}} \left( \hat c_j^\dagger \hat c_{j+L/2} + \hat c_{j+L/2}^\dagger \hat c_j \right) \,, \label{TL2}
\end{eqnarray}
where we assumed in Eq.~(\ref{TL2}) that $L$  is an even integer.
A complete set of $L$ conserved one-body operators (for finite systems under considerations also called a complete set of LIOMs) is obtained by setting $n = 1,2,...,L/2-1$.

\end{document}